\documentclass[11pt,twoside]{article}

\usepackage{asp2006_hotcool}
\usepackage{graphicx}
\usepackage{epsf}
\usepackage{psfig}
\usepackage{lscape}

\begin{document}

\title{Improving stellar parameter and abundance determinations of early B-type stars}   
\author{Mar\'ia-Fernanda Nieva$^{1}$ \& Norbert Przybilla$^{2}$}   
\affil{
$^{1}$ MPI for Astrophysics, Postfach 1317, 85741 Garching, Germany\\
$^{2}$ Dr. Remeis Observatory, Sternwartstr. 7, 96049 Bamberg, Germany}    

\begin{abstract} 
In the past years we have made great efforts to reduce the statistical
and systematic uncertainties in stellar parameter and chemical
abundance determinations of early B-type stars. Both the construction of 
robust model atoms for non-LTE line-formation
calculations and a novel self-consistent spectral analysis methodology 
were decisive to achieve results of unprecedented precision.
They were extensively tested and
applied to high-quality spectra of stars from OB associations and the
field in the solar neighbourhood, covering a broad parameter range.
Initially, most lines of hydrogen, helium and carbon in the optical/near-IR spectral range 
were reproduced simultaneously in a consistent way for the first time,
improving drastically on the accuracy of results in published work.
By taking additional ionization equilibria of oxygen, neon, silicon and iron into 
account, uncertainties as low as
$\sim$1\% in effective temperature, $\sim$10\% in surface gravity and
$\sim$20\% in elemental abundances are achieved -- compared to $\sim$5-10\%,
$\sim$25\% and a factor $\sim$2-3 using standard~methods.

Several sources of systematic errors have been identified when comparing our
methods for early B-type stars with standard techniques used in the nineties 
and also recently  
(e.g. VLT-FLAMES survey of massive stars). 
Improvements in automatic analyses are strongly recommended for meaningful comparisons
of spectroscopic stellar parameters and chemical abundances ('observational constrains')
with predictions of stellar and galactochemical evolution models.
\end{abstract}

\section{Introduction}
Normal unevolved early OB-type stars of $\sim$8-20 M$_\odot$ are the
objects with the simplest photospheric physics among the massive stars. 
They are unaffected by e.g. strong stellar winds like the hotter
and more luminous
stars or by convection and chromospheres like the cool supergiants. However,
their spectral analysis turned out to provide inconclusive results in
the past decades, i.e. too large uncertainties in basic stellar parameters
and an overall enormous range in derived elemental abundances, 
posing a challenge to predictions of stellar and Galactochemical 
evolution models (see  review by Przybilla~2008).

In order to improve the quantitative analysis of these stars 
we have exhaustively updated the spectral modelling by
constructructing robust model atoms for non-LTE line-formation
calculations. In parallel, we have implemented a powerful
self-consistent analysis technique, which brings numerous
spectroscopic parameter and abundance indicators into agreement simultaneously.

Our efforts have provided highly-promising results so far, i.e. a drastic
reduction of statistical and systematic uncertainties in stellar
parameters and chemical abundances (Nieva \& Przybilla~2007, 2008; NP07/08). 
As a first application, 
stars from OB associations and the field in the solar neighbourhood
covering a broad parameter range were analysed. The sample turned out
to be chemically homogeneous on the $\sim$10\% level (Przybilla, Nieva
\& Butler 2008, PNB08), corroborating earlier findings from analyses 
of the ISM gas-phase. The data are also consistent with published 
Orion nebula abundances. The results have an
immediate impact on several fields of contemporary astrophysics like
stellar (see Przybilla, Firnstein \& Nieva, these proceedings) 
and Galactic chemical evolution models and the dust-phase
composition of the local ISM. They provide an independent view on the
discussion of photospheric solar abundances and helioseismic
constraints on the solar interior model, and they define the initial chemical 
composition for models of star and planet formation in the solar neighbourhood.
In addition, a few B-type hyper-velocity stars were analysed using this 
technique, providing valuable constraints on their nature and 
their ejection mechanisms (Przybilla et al.~2008a,b). 

Despite this kind of star has relatively simple photospheres
compared to other objects, the spectral analysis is still
sensitive to many potential systematic effects which are usually
underestimated. We discuss
the most common sources of systematic error that have to be avoided when high
precision/accuracy in spectral analyses is desired.
Observational constraints as obtained from automatic spectral analyses of large star 
samples like the VLT-FLAMES survey of massive stars may benefit significantly 
from a proper elimination of these systematics.

\section{Model calculations and new spectral analysis methodology}
A hybrid approach is used for the non-LTE line-formation
computations. These are based on line-blanketed plane-parallel,
homogeneous and hydrostatic LTE model atmospheres calculated with
{\sc Atlas9}. Non-LTE synthetic spectra are computed with recent versions of
{\sc Detail} and {\sc Surface}. These codes solve the coupled radiative transfer
and statistical equilibrium equations and compute synthetic spectra
using refined line-broadening data, respectively. 
The hybrid non-LTE approach is consistent with full non-LTE calculations
(NP07) but faster and it also allows comprehensive
model atoms based on critically selected atomic data to be employed
in the non-LTE line-formation computations. 

The new spectral analysis was originally based on a
self-consistent and simultaneous reproduction of almost all hydrogen, helium and 
carbon lines in the optical/near-IR spectra, matching multiple ionization equilibria 
(He\,{\sc i/ii},  C\,{\sc ii/iii/iv}), see Nieva \& Przybilla~(2006), NP07
and NP08 for details. The method was further extended by consideration of
additional ionization equilibria (O\,{\sc i/ii}, Ne\,{\sc i/ii}, 
Si\,{\sc ii/iii/iv}, Fe\,{\sc ii/iii}, PNB08).  
This allows unprecedently accurate stellar parameters
and elemental abundances to be derived, with uncertainties as low as
$\sim$1\% in effective temperature $T_{\rm eff}$, $\sim$10\% in surface
gravity $g$ and 
$\sim$20\% in elemental abundances. Significant improvements on
results from previous studies are thus achieved, which typically give uncertainties of 
$\sim$5-10\%, $\sim$25\% and a factor $\sim$2-3 for these quantities.
Moreover, the successful implementation of the new method required an
identification of sources of systematic uncertainties in standard spectral 
analyses and allowed in most cases for their quantification, which we
discuss in the following.

\section{Reducing systematic uncertainties}
Every step in a  quantitative spectral analysis is susceptible to systematic
uncertainties. Here we briefly list the most common sources of
systematics that affect the final error in stellar parameters and elemental abundances 
of normal unevolved early B-type stars. When possible, recipies are given how to 
prevent them. Additional systematics may arise from further complications like 
magnetic fields, but this is beyond the present scope.\\[2mm]
{\bf Single or double?}
Not only spectroscopic but also close visual binaries are expected to be observed
in dense fields. Standard analyses for single stars 
applied to a spectrum contaminated with light of a second star can give
erroneous results throughout.  
Inspection of H or He lines can help to identify asymmetries due to a companion before 
carring out an automatised quantitative analysis. \\[2mm]
{\bf Quality of spectra.}
Continuum normalization, local continuum definition and low S/N are important sources of 
systematics. E.g., spectra of S/N$\sim$50 challenge the definition of the local
continuum of spectral lines. The abundance determination in stars rotating
at intermediate velocities ($v\sin i\sim 50\,\mathrm{km\,s^{-1}}$) is
limited to 0.2-0.3\,dex in accuracy at this S/N. Fast-rotating stars ($v\sin i \ge
150\,\mathrm{km\,s^{-1}}$) or low-resolution spectra impose even more complications 
because metal line blends lower the real continuum, hence the
abundances can be systematically underestimated and the accuracy is limited to 0.3-0.4\,dex
(see Korn et al.~2005).\\[2mm]
{\bf Model atmospheres and line formation.} 
Atmospheric structures computed with {\it full} non-LTE or {\it hybrid} non-LTE
methods are equivalent for unevolved B stars (NP07), when the abundances
used for opacity calculations are the same -- note that `solar' abundances
have changed over time. {\em Line-blanketing} effects on the atmosphere impact 
the line-formation calculations, introducing dependencies on metallicity and
microturbulence. In a similar way, {\em line blocking} affects the
strength of synthetic lines by modifying radiative rates. Both effects need to be
accounted for in a consistent way.\\[2mm]
{\bf Model atoms.} Most model atoms for non-LTE calculations are
based on input atomic data ({\em ab-initio} and approximation data) as available 
in the early nineties. It is worthwhile to check for
improvements on the modelling whenever new data becomes available.
E.g., for C different model atoms yield discrepancies in abundances up to
0.8\,dex for some lines while no discrepancies are found for others, 
see Sect.~\ref{systematics} and NP08. Other elements also show a similar behaviour.\\[2mm]
\begin{figure}[!ht]
\includegraphics[width=0.43\linewidth,height=5.5cm]{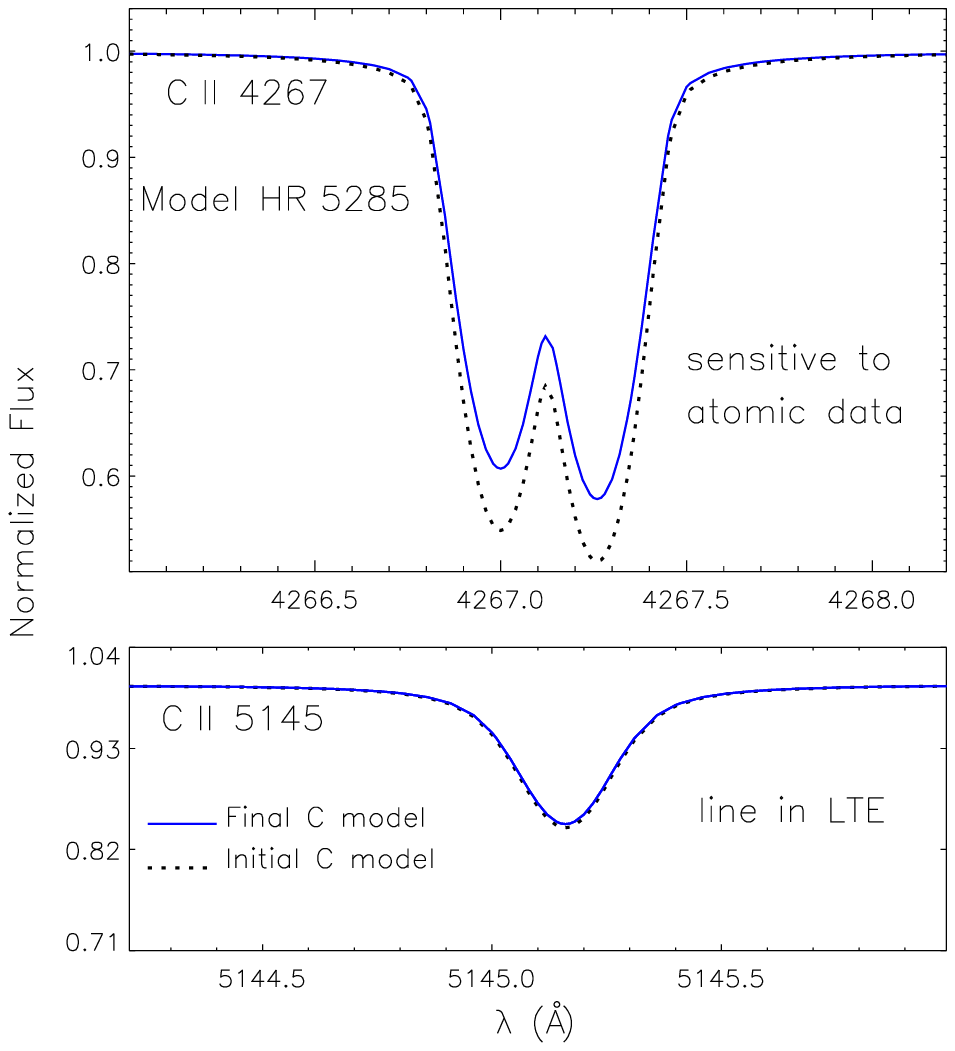}
\hfill
\includegraphics[width=0.54\linewidth,height=5cm]{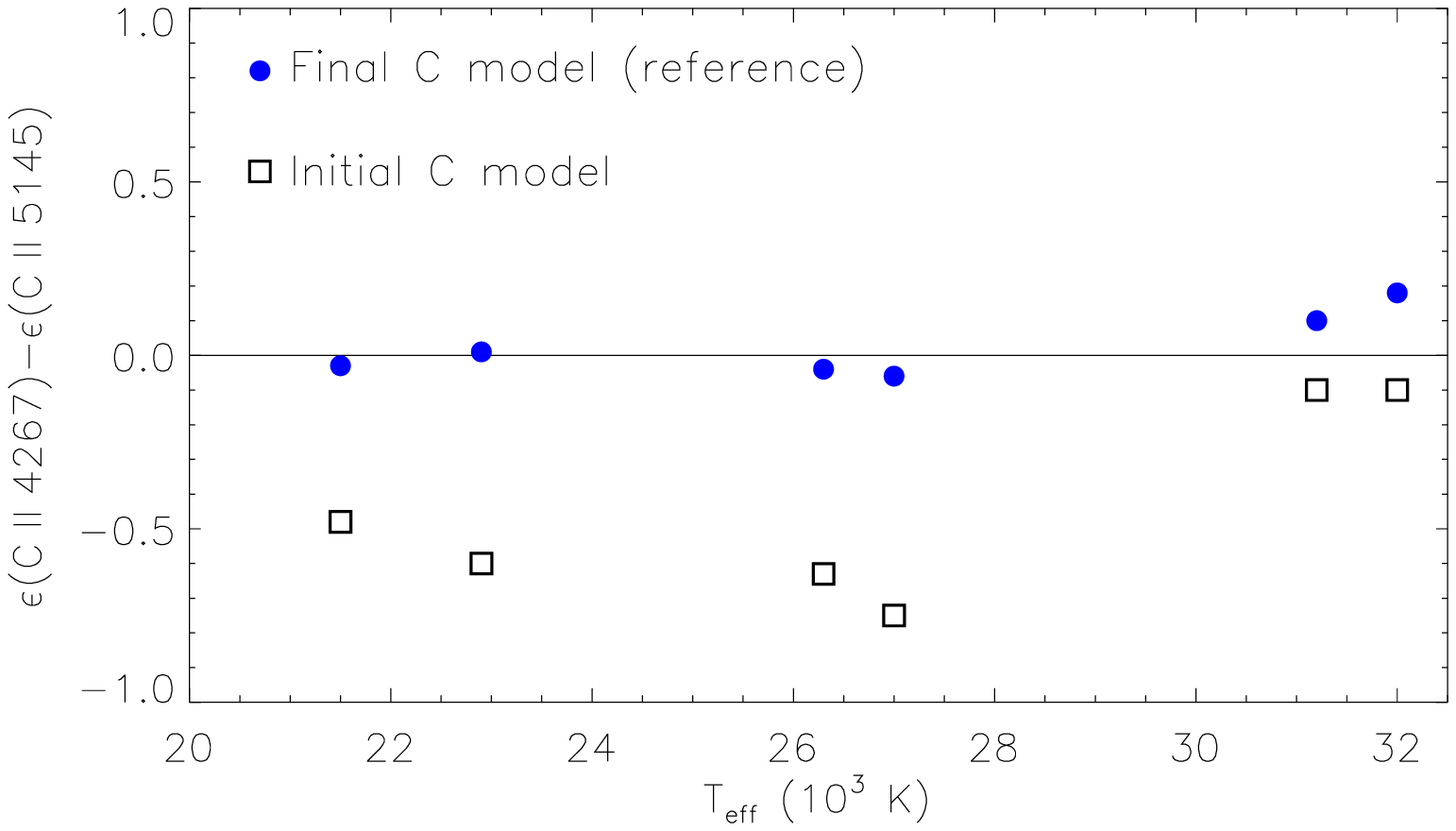}
\caption{Left: sensitivity of two C\,{\sc ii} lines
to use of photoionization cross-sections from different {\em ab-initio}
calculations in a model spectrum for a B2\,V star.
Right: differences in C abundance as derived from
the two lines in 6 stars as a function of $T_{\rm eff}$. From NP08, see text for details.}
\label{atomicdata}
\end{figure}
{\bf Effective temperatures.} $T_\mathrm{eff}$ estimated from photometry can differ 
from those determined with the self-consistent 
spectroscopic method by more than 10\% (NP08). 
Spectroscopic determinations via ionization equilibria
are a powerful technique only when the model atoms are reliable. 
In addition, {\em one} ionization equilibrium alone does not provide accurate
constraints because of dependencies on other variables like microturbulence  
(see below, Fig.~\ref{si-ie} and consequences for C abundances in Table~\ref{param_abund}).
Therefore, only the simultaneous use of multiple ionization equilibria
provides a reliable and parameter-free $T_\mathrm{eff}$ determination.\\[2mm]
{\bf Surface gravity.} The common use of {\it only  one}
Balmer line as $\log\,g$  indicator does not allow for consistency checks. 
Instead, most Balmer lines and metal ionization equilibria should be
considered for an accurate determination. Moreover, 
neglecting non-LTE effects also leads to systematic errors: increasing with
$T_\mathrm{eff}$ up to $\sim$0.2\,dex in $\log\,g$ around 35\,000\,K, see
NP07/08 and Sect.~\ref{systematics} 
Ionization equilibria can also be a powerful tool for the $\log g$ determination.\\[2mm]
{\bf Microturbulence.} This quantity is generally derived from selected
lines of {\em one} oxygen or {\em one} silicon ion. However, the microturbulent
velocity $\xi$ should adopt the same value for all metals. In particular,
cross-checks including different species are mandatory 
when only a few lines are measurable (like in fast rotators) in order to
avoid large uncertainties.
Microturbulent velocities often come  close to or exceed the sound speed in 
many previous studies. In contrast, physically more reasonable lower values are
found in our work, increasing from $\sim$2-4\,${\rm km\,s}^{-1}$ in dwarfs to
$\sim$5-8\,${\rm km\,s}^{-1}$ in giants. Consequences of overestimated
$\xi$ for the $T_\mathrm{eff}$ and abundance determination are discussed in
Sect.~\ref{systematics}\\[2mm]
{\bf Spectral line selection.} Abundances may depend on the selection of lines used 
in the analysis when the model atoms are not comprehensive: some multiplets may indicate
systematically different abundances than others. Moreover, some
lines are observable only in cooler or in hotter stars. Which lines
should one choose for the analysis? All possible observed lines for each star should 
be taken into account in the optimum case  
(see NP08 and PNB08).\\[2mm]
{\bf Non-LTE corrections.} Non-LTE effects cannot be easily predicted.
They affect different lines of the same star in different ways. 
Non-LTE line strengthening or weakening can occur, and non-LTE effects are not 
restricted to stronger lines alone. For different plasma conditions the non-LTE effects 
change. Hence, adding or subtracting fixed `non-LTE abundance corrections'
to LTE results may increase the systematics. \\[2mm]
{\bf Macroturbulence.} Macroturbulence is not considered in many studies.
This is important for proper $v\sin i$ determinations in apparently
slow-rotating objects.\\[2mm]

\subsection{Systematics from atomic data and atmospheric parameters}\label{systematics}
Here we provide a few quantitative examples of systematic effects on
abundance analyses to be expected due to use of different
input atomic data for non-LTE calculations and due to
atmospheric parameter variations.

Fig.~\ref{atomicdata} shows discrepant results for C\,{\sc ii} lines predicted by two
different non-LTE model atoms. On the left panel the sensitivity of two lines 
to different {\em ab-initio} photoionization cross-sections is shown.
Inaccurate atomic data will indicate largely underestimated abundances 
for the stronger line. This is quantified for stars within a broad parameter 
range (right panel). The discrepancies in abundances depend on the reliability of the 
model atom and on the plasma conditions of the star and amount up to 0.8\,dex. 
In parallel, collisional excitation and ionization
cross-sections and $\log\,gf$ values need to be reliable (NP08). 

\begin{figure}[!t]
\includegraphics[width=0.95\linewidth]{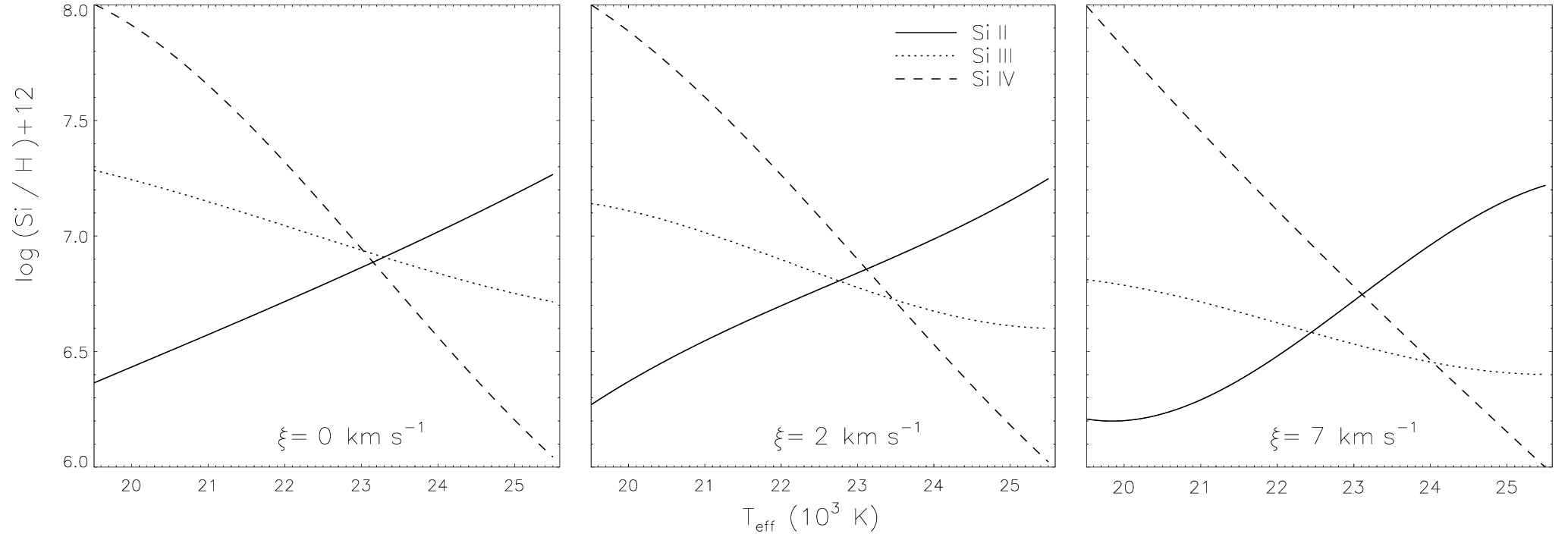}
\caption{Effective temperature determination via Si ionization equilibrium for 
the LMC star NGC2004-D15.
The Si\,{\sc iii} lines are strong, hence the
derived Si\,{\sc iii} abundance depends on the microturbulence $\xi$. Different
$T_\mathrm{eff}$ is derived from Si\,{\sc ii/iii} and Si\,{\sc iii/iv}
for ill-chosen $\xi$, e.g. an overestimation of $\xi$ by 7\,km\,s$^{-1}$ gives $\Delta
T_\mathrm{eff}$\,$\approx$\,2000\,K. The Si\,{\sc ii/iii/iv} ionization
equilibrium is established only for the correct $\xi$ (from Nieva~2007).}
\label{si-ie}
\end{figure}

\begin{table}[!ht]
\caption[]{Systematic errors in C
abundances of  individual  lines 
caused by atmospheric parameter variations and the assumption of
LTE for the line-formation calculations in the B1\,III star HR\,3055
 ($T_\mathrm{eff}=31\,200 \pm 300$\,K; $\log g= 3.95 \pm 0.05$; $\xi=
8 \pm 2\,\rm{km\,s^{-1}}$, see NP08, PNB08)}
\label{param_abund}
 $$
\footnotesize{
 \begin{array}{lcccccccc}
 \noalign{}
  \hline
 \mathrm{Ion} & \lambda & W_{\lambda} &
\mathrm{\varepsilon(C)_\mathrm{NLTE}} & \Delta T_\mathrm{eff} & \Delta\log\,g & \Delta\,\xi  & \mathrm{LTE} \\
  &  \mathrm{(\AA)}  & \mathrm{(m{\AA})}  &  & -2000\,\mathrm{K} & +0.2\,\mathrm{dex}& +5\,\mathrm{km\,s}^{-1}& \\
 \hline\\[-2mm]
\rm C$\,{\sc ii}$ &4267.2 & 113 & 8.46 &-0.33 &-0.11 &-0.16  &-0.40  \\[-.5mm]
           &5133.3 & 19 & 8.34 &-0.30 &-0.10 &~~0.00 &~~0.00  \\[-.5mm]
           &5143.4 & 10 & 8.44 &-0.40 &-0.05 &~~0.00 &~~0.00  \\[-.5mm]
           &5145.2 & 19 & 8.36 &-0.32 &-0.09 &-0.02  &~~0.00  \\[-.5mm]
           &5151.1 & 11 & 8.34 &-0.30 &-0.08 &~~0.00 &~~0.00  \\[-.5mm]
           &5662.5 & ~4 & 8.37 &-0.33 &-0.13 &~~0.00 &~~0.00  \\[-.5mm]
           &6578.0 & 66 & 8.24 &-0.40 &-0.15 &-0.10  &~~0.00  \\[-.5mm]
           &6582.9 & 37 & 8.29 &-0.30 &+0.02 &+0.05  &+0.05  \\[1mm]
\rm C$\,{\sc iii}$ &4056.1& 45 & 8.28 &+0.21 &+0.06 &-0.04  &+0.08   \\[-.5mm]
             &4162.9 & 58 & 8.29&+0.28 &+0.09 &-0.03  &+0.25   \\[-.5mm]
             &4186.9 & 92 & 8.34&+0.35 &+0.15 &-0.08  &+0.07   \\[-.5mm]
             &4663.5 & 18 & 8.27 &+0.22 &+0.07 &-0.03  &+0.22   \\[-.5mm]
             &4665.9 & 50 & 8.24 &+0.26 &+0.08 &-0.08  &+0.35   \\[-.5mm]
             &5272.5 & 14 & 8.28 &+0.16 &+0.01 &~~0.00 &~~0.00   \\[1mm]
\rm C$\,{\sc iv}$&5801.3 & 53 & 8.45& +1.06 &+0.46& -0.03  &+0.39\\[-.5mm]
          &5811.9 & 34 & 8.45 & +1.06 &+0.46& -0.03  &+0.39   \\[-.5mm]
 \hline
 \end{array}
}
 $$
 \end{table}

Fig.~\ref{si-ie} shows the dependency of $T_\mathrm{eff}$ to the adopted
value of microturbulence when only one ionization equilibrium of Si is 
adopted, i.e. Si\,{\sc ii/iii} or Si\,{\sc iii/iv}. This will also
affect the derivation of $\log\,g$ from the Balmer lines. This problem can
be solved via use of multiple ionization equilibria, e.g. Si\,{\sc
ii/iii/iv} and when possible also considering other elements (as
explained above). The goal should be deriving the same value of $T_\mathrm{eff}$, 
$\log\,g$ and $\xi$ from all H, He and all metals.  

Systematic errors in C abundances 
from individual lines in a B1\,III star
due to variations in $T_\mathrm{eff}$, $\log\,g$ and
$\xi$ have been quantified in Table~\ref{param_abund}. The offsets in parameters are averaged
discrepancies for studies using standard analysis techniques.
Incorrect parameters prevent consistent C\,{\sc ii/iii/iv}
ionization equilibria to be achieved, even when the model atom is highly reliable
in the defined parameter range.
Parameters derived from C ionization equilibrium
(NP08) are confirmed by other metals (PNB08) when reliable model atoms are~used.

\section{Conclusions}
In the past years we have made great efforts to reduce uncertainties
in quantitative spectral analyses of early B-type stars. 
A self-consistent analysis, i.e. account of all spectroscopic
indicators -- Balmer and helium lines and multiple metal ionization equilibria
-- throughout the optical and near-IR, resulted in drastically reduced systematic 
effects in the atmospheric parameter and elemental abundance determination.
A large number of potetial systematic errors was identified when comparing our 
new models and self-consistent method with standard techniques.
One should keep in mind that statistics does by no mean reduce systematic errors. 
We conclude that careful improvements in automatic spectral analysis routines
should be implemented before being applied to large samples of stellar spectra. 
This will prevent unnecessary systematic bias in
stellar parameters and abundance determinations, inconclusive 
results and misinterpretations of the studies by theoreticians.
Comparisons of 'observational constraints' and theoretical predictions 
are only meaningful when the former are unbiased by systematic error.

\acknowledgements 
The authors thank D.\,J. Lennon and D.\,J. Hillier for fruitful discussions.

\end{document}